\documentclass[graybox]{svmult}

\usepackage{type1cm}        
\usepackage{makeidx}         
\usepackage{graphicx}        
\usepackage{multicol}        
\usepackage[bottom]{footmisc}

\usepackage{newtxtext}       %
\usepackage[varvw]{newtxmath}       

\makeindex             

\newcommand{\supp}{\mathrm{supp}\, }               
\usepackage[sort,nocompress]{cite}
\begin{document}

\title*{Quantum Systems at The Brink}
\author{Dirk Hundertmark \and Michal Jex \and Markus Lange}
\institute{Dirk Hundertmark
\at Department of Mathematics, Institute for Analysis, Karlsruhe Institute of Technology, 76128 Karlsruhe, Germany and Department of Mathematics, Altgeld Hall, University of Illinois at Urbana-Champaign, 1409 W. Green Street, Urbana, IL 61801, USA\\
\email{dirk.hundertmark@kit.edu}
\and
Michal Jex
\at Department of Physics, Faculty of Nuclear Sciences and Physical Engineering, Czech Technical University in Prague, B\v rehov\'a 7, 11519 Prague, Czech Republic and CEREMADE, Universit\'e Paris-Dauphine, PSL Research University, Place de Lattre de Tassigny, 75016 Paris, France\\ 
\email{michal.jex@fjfi.cvut.cz}
\and
Markus Lange
\at Mathematics Area, Scuola Internazionale Superiore di Studi Avanzati (SISSA), via Bonomea 265, 34136 Trieste, Italy\\
\email{mlange@sissa.it}}
\maketitle

\abstract*{We present a method to calculate the asymptotic behavior of eigenfunctions of Schr\"odinger operators that also works at the threshold of the essential spectrum. It can be viewed as a higher order correction to the well-known WKB method which does need a safety distance to the essential spectrum. We illustrate its usefulness on examples of  quantum particles in a potential well with a long-range repulsive term outside the well.}

\abstract{We present a method to calculate the asymptotic behavior of eigenfunctions of Schr\"odinger operators that also works at the threshold of the essential spectrum. It can be viewed as a higher order correction to the well-known WKB method which does need a safety distance to the essential spectrum. We illustrate its usefulness on examples of  quantum particles in a potential well with a long-range repulsive term outside the well.}

\section{Introduction}
Except for the famous Wigner-von Neumann potentials \cite{vonNeuWig29}, bound 
states of quantum systems are usually found below the energies of scattering states. 
The bound state energies and the scattering energies are separated 
by the ionization threshold corresponding to the essential spectrum 
threshold. Above this threshold, the particles cease to be bound and 
move to infinity. Below the threshold, the binding energy, i.e., the
difference between the ionization threshold and the energy of the bound 
state, is positive. 
Since the minimal energy cost to move a particle to infinity is 
given by the binding energy and since regular perturbation theory predicts 
that the energy changes only little under small 
perturbations the quantum system is 
stable under small perturbations.  As long as the binding energy 
stays positive the corresponding eigenfunctions are still bound, 
i.e., they do not  suddenly disappear.
 
Imagine a parameter of the quantum system being tuned such that the energy of a bound 
state, e.g., the ground state energy, approaches the ionization 
threshold. 
At this critical value, the perturbation theory in the parameter 
breaks down. Moreover, at this threshold there is no energy penalty for moving the quantum particle to infinity anymore. 
So  it is unclear what happens \emph{exactly at} this 
binding--unbinding transition: Does the bound state disappear, i.e., 
the quantum particle can move to infinity and the eigenstate of the 
quantum system spreads out more and more and dissolves, or does the 
bound state still exist at the critical parameter and then suddenly 
disappears (see, for example, the discussion in \cite{KlSi-1}).
Consider a Schr\"odinger operators of the form
\begin{eqnarray}
\label{eq:intro}
	H_\lambda = -\frac{1}{2m}\Delta-V_\lambda(x) + U(x) 
\end{eqnarray}
where $ -\frac{1}{2m}\Delta$ is the kinetic energy, $U$ a non-zero repulsive part of the potential and $-V_\lambda$  a compactly supported attractive part of the potential depending on a parameter $\lambda$. This operator describes one-particle models, however with slight modifications it can also describe interacting many-particle systems.
The well-known WKB asymptotics, see also the work of Agmon\cite{Agm82}, shows that the eigenfunction $\psi_\lambda$
corresponding to a discrete eigenvalue $E_\lambda$ of the operator \eqref{eq:intro} falls off exponentially with the distance to the origin, i.e.,
\begin{eqnarray*}
\psi_\lambda \sim\exp\left(-\sqrt{2m\Delta E_\lambda}|x|\right) 
\end{eqnarray*}
for $|x| \to \infty$ where $\Delta E_\lambda \geq 0$ is the binding 
energy, i.e., the distance of the eigenvalue $E_\lambda$ to the bottom of 
the essential spectrum of $H_\lambda$. 
Such a decay estimate \emph{does not provide any useful information 
at critical coupling} when $\Delta E_\lambda=0$. 
Even worse, all 
rigorous approaches for decay estimates of eigenfunctions usually 
provide upper bounds of the form    
\begin{align}\label{eq:upper bound with loss of delta}
	|\psi(x)|\le C_\delta \exp\left(-(\sqrt{2m\Delta E_\lambda}-\delta)|x|\right) 
\end{align}
for all small enough $\delta>0$ with a constant $C_\delta$ which diverges 
in the limit $\delta\to 0$, see e.g. \cite{Agm82}.
Therefore, in order to be able to prove bounds on the asymptotic behavior 
of bound states, which still yields useful information when the binding 
energy vanishes,  
a new approach is needed. It can not require a gap 
between the eigenvalue and the threshold of the essential spectrum to work. 

\begin{figure}
\centering
\includegraphics[width=.6\columnwidth]{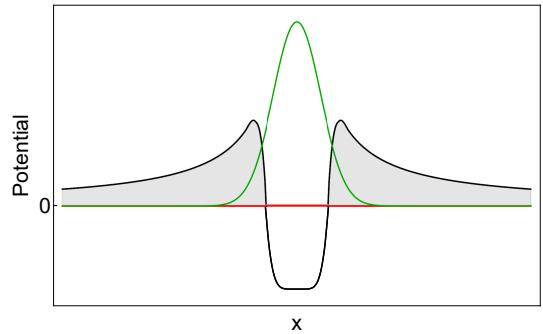}
\caption{Sketch of tunneling problem for the ground state at zero energy:  black line corresponds to the potential, red line to the energy level, green to the eigenstate and grey area to classically forbidden region.}
\vspace{-0.4cm}
 \label{fig:tunneling}
\end{figure}

The new method developed in \cite{HunJexLan21-Helium}, which is presented 
in the next section, can be viewed as a higher order correction to the WKB 
method. The main ingredient is still a suitable energy estimate. 
However, our approach to energy estimates is based on the idea 
that a \emph{positive long--range repulsive part} of the potential can 
stabilize a quantum system. Such a long range positive part allows us 
to gain  extra flexibility, in particular, it remove the necessity 
of positive binding energy, i.e., a safety distance with respect to 
the bottom of the essential spectrum. 
The underlying intuition is that if the binding energy 
$\Delta E_\lambda$ vanishes as the parameter $\lambda$ 
approaches a critical value, the bound state can only disappear when 
it tunnels through the positive tail of potential, see Figure~\ref{fig:tunneling}.
If this tunneling probability is zero, the ground state cannot disappear, hence the quantum system stays bounded at the critical coupling.  This behavior is also predicted by numerical calculations 
\cite{BusDraEstMoi14,DubIva98,Hog95,KaiSer01}. 
Our method makes this intuition precise, including upper 
bounds on the asymptotic behavior of the corresponding eigenfunctions at the ionization threshold.  

Before we present our approach let us shortly mention some known results for the existence and non-existence of threshold eigenvalues. 
Early results on existence or non-existence of threshold eigenvalues go back to
 \cite{JeKa79, Agm70, Ken89, Kno78-2, Kno78-1, Lie81, New77,Ram87, Ram88, Sim81,Yaf75}. 
In \cite{bol85} it was noted that a long range Coulomb part can create zero energy eigenstates, see also \cite{Nak94, Yaf82}. 
An analysis of eigenstates and resonances at the threshold for the case of certain nonlocal operators recently appeared in \cite{KaLo20}. 
The references presented above are by no means exhaustive. 

The main result of the paper, presented in Theorem \ref{thm:one}, 
yields decay estimates for bound states of quantum systems which 
\emph{ do not require} that the binding energy $\Delta E_\lambda$ 
is positive if a suitable long-range repulsive part of the potential 
is present.

\section{The Method}
\label{method}
For simplicity of the exposition we will only consider 
one--particle Schr\"odinger operators $H_\lambda$ of the 
form \eqref{eq:intro}
in the following. We also assume that the potentials $V_\lambda$ and 
the long range repulsive part $U$ are in the Kato--class, see \cite{AizSim82, CycFroKirSim87} or \cite{Sim82}  for the definition. 
This ensures that the potentials are infinitesimally form bounded 
with respect to the kinetic energy $-\Delta$, so the Schr\"odinger operator 
\eqref{eq:intro} is well--defined with the help of quadratic form methods, 
\cite{ReSi4, Tes14}. 
The ionisation threshold 
$\Sigma_\lambda= \inf\sigma_{\rm{ess}}(H_\lambda)$ is given by the bottom of 
the essential spectrum.  
We also assume that the potential vanishes at infinity, 
in which case $\sigma_{\text{ess}}(H_\lambda)= [0,\infty)$, 
i.e., $\Sigma_\lambda= 0$.   
\begin{theorem} \label{thm:one}
Each normalized eigenfunction $\psi_\lambda$ corresponding to an eigenenergy $E_\lambda \leq 0$ of $H_\lambda$ satisfies 
\begin{eqnarray}\label{eq:AbstractDecayBound}
	|\psi_\lambda| \lesssim \exp\left({-F - \frac{1}{2}\ln{\left(\Delta E_\lambda+ U - \frac{|\nabla F|^2}{2m}\right)}}\right)
\end{eqnarray}
with $\Delta E_\lambda=-E_\lambda$ being the binding energy, 
and $F$ being any function which is bounded from below and satisfies  
\begin{eqnarray}\label{eq:ConditionOnF}
	\frac{|\nabla F|^2}{2m} < \Delta E_\lambda+ U
\end{eqnarray}
for all $|x|\geq R>0$. 
Here $U\ge 0$ is the repulsive part of the potential.  
\end{theorem} 
\begin{remark}\label{rem:two}
Choosing  $F(x)=\mu|x|$ yields $|\nabla F(x)|^2 =\mu^2$. 
Note that  $\Delta E_\lambda+ U - \frac{|\nabla F|^2}{2m}\ge 
\Delta E_\lambda  - \frac{\mu^2}{2m}$ since $U\ge 0$.  
Thus in the subcritical case, when the binding energy is positive, 
upper bounds of the form \eqref{eq:upper bound with loss of delta}, 
which, for one--particle operators, coincide with the result of Agmon 
\cite{Agm82},  follow immediately from Theorem \ref{thm:one}. 	

In the critical case, when the binding energy vanishes, a 
non-zero repulsive part $U$ is indispensable since 
otherwise \eqref{eq:ConditionOnF} can never be satisfied. However, 
in contrast to the usual WKB asymptotics our bound provides 
detailed information on how well the quantum system is localized 
at critical coupling, when a repulsive part $U$ is present. 
The logarithmic expression in the exponent of 
\eqref{eq:AbstractDecayBound} corresponds to a polynomial correction of 
the asymptotic behavior and in all relevant cases it is of smaller order 
than $F$.

The existence of the  eigenstate 
is a necessary assumption 
in Theorem~\ref{thm:one}. On the other hand, as shown 
in \cite{HunJexLan21-Helium,HunJexLan22-Potential}, the 
existence of an eigenstate 
for the critical case follows from bounds of the form 
\eqref{eq:AbstractDecayBound} together with tightness 
arguments in the form of, e.g., \cite{HunLee12}.  
\end{remark}

\begin{proof}
  In the following we will, for notational simplicity, drop the 
  dependence of the Schr\"odinger operator, the wave function, and the 
  eigenenergy on the parameter $\lambda$. 

\smallskip 
\noindent 
\textbf{Starting point:} 
Consider a self-adjoint operator $H$ given in \eqref{eq:intro} with and 
a normalized eigenvector $\psi$ satisfying
\begin{eqnarray*}
	H\psi=E\psi\,
\end{eqnarray*}
where $E$ is the corresponding eigenvalue below or at the threshold of the 
essential spectrum.

\noindent
\textbf{1st step:} 
Let $0\le \chi\le 1$ be a  smooth real--valued function satisfying 
\begin{eqnarray}
	\chi(x) = \begin{cases}
		0\,, \; \textrm{for } |x|\le 1 \\
		1\,,\; \textrm{for } |x|\ge 2
	\end{cases} \; . 
\end{eqnarray}
The scaled functions given by  $\chi_R(x)= \chi(x/R)$ 
for $R>0$ smoothly localize in the region $\{|x|\ge R\}$. 
Note that $\mathrm{supp} \nabla \chi_R $ is localized in the annulus 
$ \{R\le |x|\le  2R  \}$. 

Let $F$ be another smooth and bounded real--valued function for 
which also $|\nabla F|$ is bounded. With  $\xi=\chi_R e^F$  
one calculates from the eigenvalue equation 
\begin{eqnarray*}
	\mathrm{Re}\langle(\xi)^2\psi,H\psi\rangle
		=  E\langle(\xi)^2\psi,\psi\rangle=E\|\xi\psi\|^2\, . 
\end{eqnarray*}

\noindent
\textbf{2nd step: }
 Using a variant \cite{CycFroKirSim87,Gri04}  of the IMS localization 
 formula \cite{Ism61,Mor79,Sig82}, we obtain 
\begin{eqnarray*}
E\|\chi_R e^F\psi\|^2\ 
  = \mathrm{Re}\langle(\xi)^2\psi,H\psi\rangle
  = \langle \xi\psi,H \xi\psi\rangle
  	-\frac{1}{2m}\langle\psi,|\nabla \xi|^2\psi\rangle\,.
\end{eqnarray*}
Clearly, $\nabla \xi = \nabla(\chi_R e^F) = \nabla\chi_r e^F + \chi_R\nabla F e^F$, so 
\begin{equation*}
  |\nabla \xi|^2 
	\le \left( 
			|\nabla\chi_R|^2
			+2 \chi_R |\nabla\chi_R||\nabla F|  
		\right) e^{2F}
			+ |\nabla F|^2\xi^2  \,.
\end{equation*}
Note that the good part $G=  \left( 
			|\nabla\chi_R|^2
			+2 \chi_R \nabla\chi_R\cdot\nabla F  
		\right) e^{2F}
$
has compact support, because the support of $\nabla\chi_R$ is compact for 
any $R>0$.   
Rearranging the terms, we obtain
\begin{eqnarray}\label{eq:geil}
	\Big\langle \chi_R e^F \psi,\Big(H-E-\frac{1}{2m}|\nabla F|^2\Big) \chi_R e^F \psi\Big\rangle\leq \frac{1}{2m}\langle\psi,G\psi\rangle \, .
\end{eqnarray}
The usual argument now uses Persson's theorem \cite{per60} for the 
bottom of the essential spectrum    
\begin{equation}\label{persson}
	\Sigma=\inf\sigma_{\text{ess}}(H) 
	=\lim_{R\to\infty}\{ \langle\varphi,H\varphi\rangle :\, 
	\|\varphi\|=1, \supp(\varphi)\subset B_R^c\}
\end{equation} 
where $B_R^c= \{|x|\ge R\}$. Thus, since we assume that $\Sigma=0$,  
for any $\delta>0$ there exist $R_\delta<\infty$ such that 
\begin{align*}
	\langle\varphi,H\varphi\rangle>(\Sigma-\delta)\langle\varphi,\varphi\rangle = -\delta\langle\varphi,\varphi\rangle 
\end{align*}
for all $\varphi$ with support outside a centered ball of 
radius $R_\delta$. So with $R=R_\delta$, we get from \eqref{eq:geil} 
\begin{align*}
		\Big\langle \chi_R e^F \psi,\Big(-\delta-E-\frac{1}{2m}|\nabla F|^2\Big) \chi_R e^F \psi\Big\rangle\leq \frac{1}{2m}\langle\psi,G\psi\rangle
\end{align*} 
but one needs positivity of 
$-\delta-E-\frac{1}{2m}|\nabla F|^2 $ 
and this requires $E<-\delta$, i.e., a safety distance of the negative 
eigenvalue to  the essential spectrum. 

Instead, we use the assumption that the potential is given by 
$-V+U$, where $V$ has compact support and $U$ 
is positive. Chosing $R$ so large that the support of $V$ is 
contained in $\{|x|\le R\}$, we have  
\begin{align*}
	\langle \chi_Re^F\psi, H\chi_Re^F\psi\rangle 
	  = \langle \chi_Re^F\psi, (-\Delta+U)\chi_Re^F\psi\rangle
	  \ge  \langle \chi_Re^F\psi, U\chi_Re^F\psi\rangle 
\end{align*}  
and using this in \eqref{eq:geil} one arrives at 
\begin{align}\label{eq:geil2}
	\Big\langle \chi_R e^F \psi,\Big(\Delta E +U-\frac{1}{2m}|\nabla F|^2\Big) \chi_R e^F \psi\Big\rangle\leq \frac{1}{2m}\langle\psi,G\psi\rangle \, .
\end{align}
where $\Delta E= -E$ is the binding energy. We want to use this energy 
inequality to prove exponential bounds on $\psi$, but for this we need 
that $F$ is growing.  

\noindent 
\textbf{3rd step:} In order to overcome the requirement that $F$ is 
bounded, we regularize it. Let $F$ be smooth, bounded 
from below and let $\nabla F$ be bounded. Adding a constant to $F$, we can assume that $F\ge 0$. 
This also does not change the gradient of $F$.
Then for any $\varepsilon>0$ the function  
\begin{align*}
	F_\varepsilon= \frac{F}{1+\varepsilon F}
\end{align*}
is smooth and bounded. Since 
$\nabla F_\varepsilon = (1+\varepsilon F)^{-2}\nabla F$ also 
$\nabla F_\varepsilon$ is bounded. 
Let $\xi_\varepsilon$ and $G_\varepsilon$ be defined 
as above with $F$ replaced by $F_\varepsilon$. 
Clearly $F_\varepsilon \le F$ and 
$|\nabla F_\varepsilon| \le  |\nabla F|$ for all 
$\varepsilon\ge 0$. Hence  $G_\varepsilon \le G$ and 
\begin{align*}
  |\nabla\xi_\varepsilon|^2 
  	\le G_\varepsilon +|\nabla F_\varepsilon|^2\xi_\varepsilon^2
  	\le G +|\nabla F|^2\xi_\varepsilon^2
\end{align*}
for all $\varepsilon \ge 0$. The argument leading to 
\eqref{eq:geil2} then shows 
\begin{align}\label{eq:geil3}
	\Big\langle \chi_R e^{F_\varepsilon} \psi,\Big(\Delta E +U-\frac{1}{2m}|\nabla F|^2\Big) \chi_R e^{F_\varepsilon} \psi\Big\rangle\leq \frac{1}{2m} \langle\psi,G\psi\rangle \le K \|\psi\|^2\, .
\end{align}
with $K=\frac{1}{2m}\sup_{R\le |x|\le 2R} G(x)<\infty$, since $G$ is supported inside  
$\{R\le |x|\le 2R\}$.  

Note that 
\begin{align*}
	\Big\langle \chi_R e^{F_\varepsilon} \psi,\Big(\Delta E +U-\frac{1}{2m}|\nabla F|^2\Big) \chi_R e^{F_\varepsilon} \psi\Big\rangle
	= \left\| \chi_R \E^{F_\varepsilon +\frac{1}{2}\ln\big(\Delta E +U-\frac{1}{2m}|\nabla F|^2\big)} \psi\right\|^2\,.
\end{align*}
The monotone convergence theorem and \eqref{eq:geil3} yield 
\begin{align*}
  & \left\| \chi_R \E^{F +\frac{1}{2}\ln\big(\Delta E +U-\frac{1}{2m}|\nabla F|^2\big)} \psi\right\|^2 \\
   &\quad\quad= \lim_{\varepsilon\to 0}
	\Big\langle \chi_R e^{F_\varepsilon} \psi,\Big(\Delta E +U-\frac{1}{2m}|\nabla F|^2\Big) \chi_R e^{F_\varepsilon} \psi\Big\rangle 
     \le K \|\psi\|^2<\infty\, .
\end{align*}
for any normalized eigenfunction $\psi$ with energy $E\le 0$. 
This proves an $L^2$ exponential bound on $\psi$, i.e., the function 
\begin{align*}
	x\mapsto \exp\left(F(x) + \frac{1}{2}\ln\big(\Delta E +U-\frac{1}{2m}|\nabla F|^2\big)\right)\psi(x) 
\end{align*} 
is in $L^2$ under the condition that all exponential weights $F$ satisfy \eqref{eq:ConditionOnF}. The claimed pointwise bound on 
$\psi $ then follows from such an $L^2$ bound using subsolution 
estimates of \cite{Sim82}, see, e.g., the discussion in \cite[Corollary 5.4]{HunJexLan21-Helium}.  
\end{proof}

\section{Examples}
In this section we consider illustrative examples of a quantum particle in a potential well with a long range Coulomb repulsion term. In the first example the tunable parameter is the depth of the potential well. We will see that the bound from Theorem~\ref{thm:one} fits very well with the explicitly calculated asymptotic behavior of the ground state of such a system. In a second example we tune the strength of the repulsion term. In the last example we illustrate that a long range repulsion term is crucial at critical coupling.   \\ 

\noindent
\textit{First example:} Let us consider 
\begin{eqnarray}
\label{eq:exam}
H_\lambda=-\Delta-\lambda\,\textrm{1}_{\{|x| \leq 1\}}+\frac{\textrm{1}_{\{|x| > 1\}}}{|x|}\,.
\end{eqnarray}
Here we chose $m=\frac 1 2$ for convenience. 
In this case $U(x)= 1/|x|$ for $|x|\ge 1$. 
It can be easily shown that there exists a critical value $\lambda_{\mathrm{cr}}$ s.t. for $\lambda>\lambda_{\mathrm{cr}}$, the Hamiltonian $H_\lambda$ has at least one bound state and for $\lambda< \lambda_{\mathrm{cr}}$ there are none. Furthermore, for this system we have $\Sigma=0$ and $\lambda_{\mathrm{cr}} \approx 0.634366
$.

Take $F(x)= 2b|x|^{1/2}$. Then $\nabla F(x)= b|x|^{-1/2}$ and 
\begin{align*}
	U(x)- |\nabla F(x)|^2 = \frac{1-b^2}{|x|}>0
\end{align*}
whenever $b^2<1$ and $|x|\ge 1$. Thus Theorem \ref{thm:one} shows 
the upper bound 
\begin{align}\label{eq:streched exp bound 1}
	|\psi_\lambda |\lesssim e^{- 2b|x|^{1/2}+ \frac{1}{2} \ln|x| }
\end{align}  
for large $|x|$ and all eigenstates with energy $E_\lambda \le 0$ and all 
$0<b<1$. This is a stretched exponential decay. 

One can make the bound tighter by choosing a more general 
radial weight function. With a slight abuse of notation, we set 
$F(x)=F(|x|)$. Then $\nabla F(x) = F'(|x|)x/|x|$ and  the borderline 
case allowed, or better, just not allowed by condition \eqref{eq:ConditionOnF} is  
\begin{align*}
	\Delta E +U(r) - |F'(r)|^2 = 0 
\end{align*}
with $r=|x|$. Hence we want to solve the equation 
$F'(r)= \sqrt{\Delta E +U(r)}$. For $a,b\ge 0$ let $F_{a,b}$ be given by 
\begin{equation}\label{eq:clever weight 1}
  \begin{split}
	F_{a,b}(r)&= \int_0^r \left(a+\frac{b}{s}\right)^{1/2}\, ds \\
	&= 		\left(a+\frac{b}{r}\right)^{1/2} r + \frac{b}{\sqrt{a}}\mathrm{arcsinh}\left[ \sqrt{\frac{a r}{b}}\right]  .
  \end{split}
\end{equation}
It is easy to check that the derivative in $r$ of the right hand 
side is given by integrand $(a+b/r)^{1/2}$. 
Splitting $U(r)= 1/|r| = \delta/r +(1-\delta)/r$, for $0<\delta<1$, suggests 
to take $a = \Delta E_\lambda$ and $b=1-\delta$. 
Theorem \ref{thm:one} then gives the upper bound  
\begin{eqnarray}\label{eq:improved streched exp bound 1}
|\psi_{\lambda}(x)|\lesssim 
	e^{-F_{\Delta E_\lambda, 1-\delta}(|x|) +\frac{1}{2}\ln |x|}\,.
\end{eqnarray}
for the ground state of $H_\lambda$ with $\Delta E_\lambda\ge 0$ 
and any $0<\delta<1$.  
In the subcritical case, where the binding energy $\Delta E_\lambda>0$, 
the first part on the right hand side  of \eqref{eq:clever weight 1}
corresponds to exponential fall-off with exponential weight 
$\sqrt{\Delta E_\lambda} |x|$ (recall that we put $m=1/2$), which 
is exactly the prediction of the WKB method,  and the second one 
is the polynomial correction since 
$\mathrm{arcsinh}[y]=\ln(y+\sqrt{y^2+1})$. 

\begin{figure}
\centering
\includegraphics[width=.95\columnwidth]{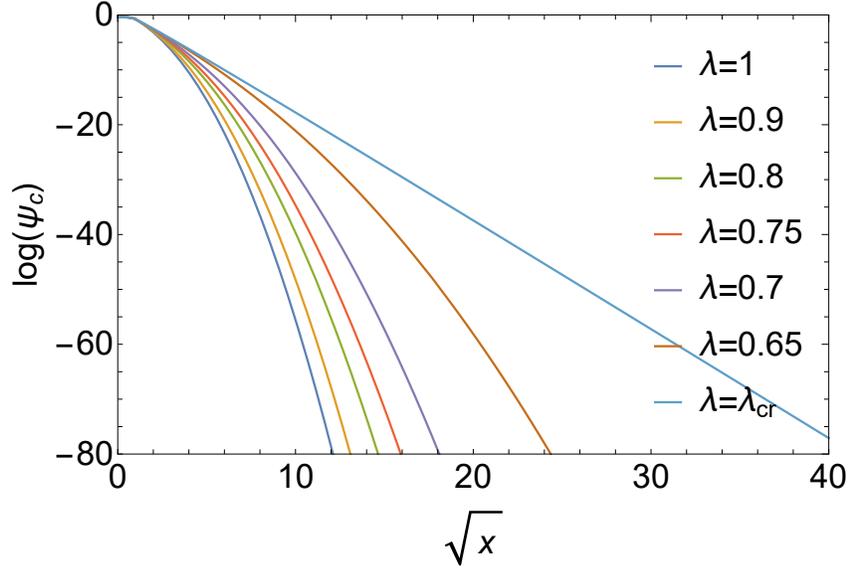}
\caption{Scaled plot of normalized ground states for the Hamiltonian \eqref{eq:exam} with varying parameter $\lambda$ for $x \in [0,1600]$. The convergence of the ground states for $\lambda \searrow \lambda_{\mathrm{cr}}\approx 0.63$ is visible. Note that in this choice of scale the parabolic curves correspond to the ground state decaying asymptotically as $\exp(-c|x|)$, as is predicted   by the WKB method, when the parameter 
$\lambda>\lambda_{\mathrm{cr}}$.  For $\lambda = \lambda_{\mathrm{cr}}$ the nearly straight line indicates that the ground state decays like $\exp(-2 \sqrt{|x|})$.}
\vspace{-0.4cm}
 \label{fig:ToyModelDecay100}
\end{figure}

Note that 
\begin{align*}
	\lim_{a\to 0} F_{a,b}(r) = 2 \sqrt{br}
\end{align*}
so in the limit where the binding energy vanishes we recover the bound 
\eqref{eq:streched exp bound 1} from \eqref{eq:improved streched exp bound 1}. 
See Figure~\ref{fig:ToyModelDecay100} for an illustration.

One can further improve upon the upper bound, by trying an 
ansatz of the form 
\begin{align}\label{eq:improved improved ansatz}
	F(r)= F_{a,b}(r)- K|x|^{\kappa} 
\end{align} 
for any $K >0$ and $0<\kappa<1/2$. It is straigtforward to check 
that with $a=\Delta E_\lambda$ and $b=1$, this ansatz  satisfies 
\eqref{eq:ConditionOnF} for all large $|x|$ and all 
$\Delta E_\lambda \ge 0$.
  
For vanishing binding energy, i.e., at $\lambda=\lambda_{\rm{cr}}$,  a matching lower bound for the ground state, which can be chosen to be strictly positive, of the form 
\begin{eqnarray*}
e^{-2\sqrt{|x|} - K |x|^\kappa}\lesssim \psi_{\lambda_{\mathrm{cr}}}(x)\, 
\end{eqnarray*}
for any $K > 0$ and $0 < \kappa < 1/2$, 
was obtained in \cite{HunJexLan21-Helium} using a subharmonic comparison lemma \cite{HofOst80-JPhysA,Agm85}. Explicit calculations show that the eigenfunction has asymptotic behavior in the form
\begin{eqnarray*}
\psi_{\lambda_{\mathrm{cr}}}(x)\sim C\frac{e^{-2\sqrt{|x|}}}{|x|^{3/4}}
\end{eqnarray*}
for large $|x|$ which is in perfect agreement with our result.

\begin{remark}
In general the existence or non--existence of ground states at critical coupling depends crucially on the dimension of the considered problem, see \cite{HunJexLan22-Potential} for more details. 
\end{remark}

\noindent
\textit{Second example:} 
We consider again an operator describing a quantum particle in a potential well with a repulsion term everywhere outside that well. 
However we do not decrease the depth of the well but increase the repulsion term. 
We start with an operator having a long range Coulomb repulsion term in three dimensions
\begin{eqnarray}
\label{eq:exam_supplementary}
H_c=-\Delta-\textrm{1}_{\{|x| \leq 1\}}(|x|)+\textrm{1}_{\{ 1<|x| \}}(|x|) \frac{c}{|x|}\,.
\end{eqnarray}

\begin{figure}
\centering
\includegraphics[width=.95\columnwidth]{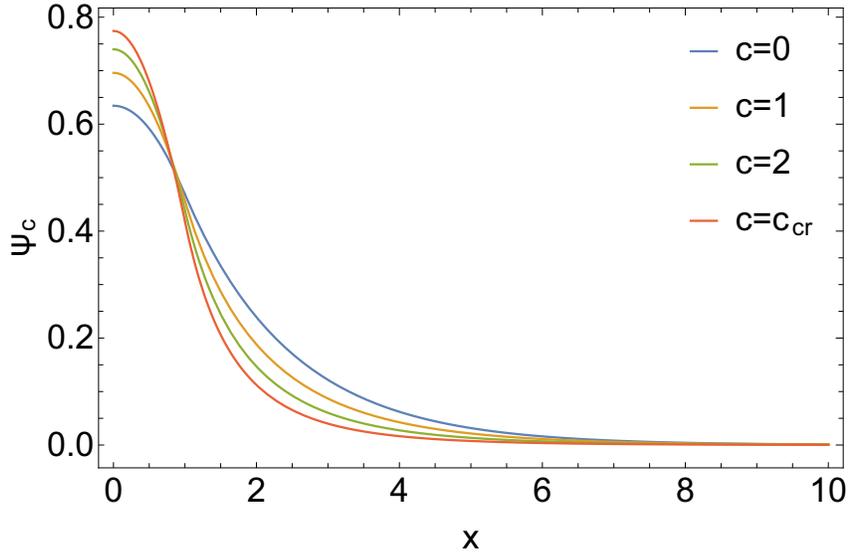}
\caption{Plot of the normalized ground state eigenfunction for the model \eqref{eq:exam_supplementary} for several values of $c$.}
\vspace{-0.3cm}
 \label{fig:ToyModel-longRangeCoulomb_supplementary}
\end{figure}

Increasing the repulsive term, i.e., increasing the parameter $c$, the eigenfunctions become more localized up to the numerically calculated critical value $c_{\mathrm{cr}} \approx 3.11693$, see Figure~\ref{fig:ToyModel-longRangeCoulomb_supplementary}. 

This operator has essential spectrum $\sigma_{\mathrm{ess}}(H)=[0,\infty)$ for any $c$ and it has negative energy ground state for sufficiently small positive $c<c_{\mathrm{cr}}$. 
The argument from Example 1 shows that eigenfunctions $\psi_c$ of $H_c$ 
with energy $E_c \le 0$ decay as  
\begin{equation*}
	|\psi_c(x)| \lesssim \exp\left( -F_{|E_c|, c}(x) +\kappa |x|^\delta \right)
\end{equation*}
for any $\kappa>0$ and $0<\delta<1/2$, where $F_{|E_c|, c}$ is given by  \eqref{eq:clever weight 1}
with the choice $a=|E_c|$ and $b=c$. 

At critical coupling $c=c_{\mathrm{cr}}$ the operator \eqref{eq:exam_supplementary} has a normalizable ground state with eigenvalue $0$. For this it is crucial to have a long range repulsive term. Without long range repulsion the eigenfunctions will delocalize more and more for $c \nearrow c_{\mathrm{cr}}$, as is illustrated in the next example.\\

\begin{figure}
\centering
\includegraphics[width=.95\columnwidth]{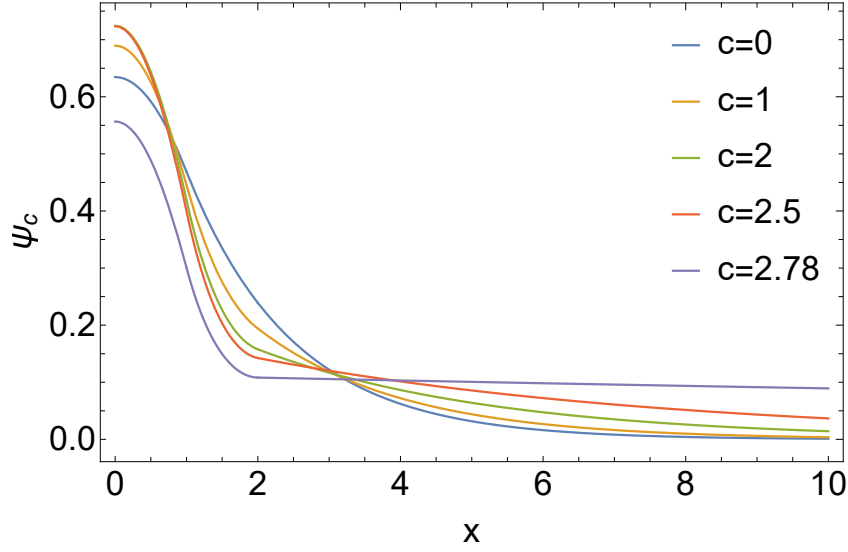}
\caption{Plot of the normalized ground state eigenfunction for the model \eqref{eq:NOlongRangeCoulomb} for several values of $c$. It illustrates, that as $c $ approaches $\widetilde{c}_{\mathrm{cr}}$ the wavefunction delocalizes. }
\label{fig:ToyModel-finitePotentialBarrier}
\vspace{-0.3cm}
\end{figure}

\noindent
\textit{Third example:} 
To see the importance of long range behavior of the repulsive potential we consider next a Hamiltonian with only a finite size repulsive barrier
\begin{eqnarray}
\label{eq:NOlongRangeCoulomb}
\widetilde{H}_{c}=-\Delta-\textrm{1}_{\{|x| \leq 1\}}(|x|)+c\textrm{1}_{\{1<|x|<2\}}(|x|)\,,
\end{eqnarray}
again in dimension three. 
Note that the value $2$ is artificial and has no particular importance. If we start to increase the parameter $c$ up to the critical value $\widetilde{c}_{\mathrm{cr}} \approx 2.7938776$, we see that far away from the critical value the increase of $c$ leads to the localization of the wavefunction even by a short range potential. However for $c \geq 2.5$ the wavefunction starts to spread further and further and for $c = 2.78$ the fall-off of the function is hardly visible, see Figure~\ref{fig:ToyModel-finitePotentialBarrier}. Using results of \cite{HunJexLan22-Potential} it is easy to see that $0$ is not an eigenvalue of the operator given in \eqref{eq:NOlongRangeCoulomb} for $c=\widetilde{c}_{\mathrm{cr}}$.\\

The presented plots highlight the physical intuition that the wavefunction has to tunnel through the repulsive barrier in order to leave the potential well and delocalize. 
However the long range Coulomb repulsion is too \emph{sticky} for the wavefunction to delocalize even at the critical value and hence we are able to prove fall-off behavior at the threshold of the essential spectrum.

\section{Outlook}
Our method is not restricted to a Coulomb type long range part of 
the potential nor to the case of one-particle models. A variety of 
physical systems can be handled. For example, it is easy to check that 
for a long range repulsive potential $U$, which is radial, say, any 
exponential weight $F$ of the form 
\begin{align}
	F(r) = \delta \int_{r_0}^r \sqrt{U(s)}\, ds  
\end{align}  
for some $r_0\ge 0$ and $0<\delta<1$ will satisfy \eqref{eq:ConditionOnF}. 
To yield a useful upper bound one need that 
$\lim_{r\to \infty} F(r)=\infty$, i.e., the integral $\int_{r_0}^r \sqrt{U(s)}\, ds$ should diverge in the limit $r\to\infty$. 
For power law repulsive potentials of the form $U(r)= c_1 r^{-\alpha}$ 
this shows that one needs  $\alpha \le 2$. Since for vanishing 
binding energy $\Delta E=0$, the correction term satisfy
$$- \frac{1}{2}\ln\left( U(r) - \frac{|\nabla F(r)|^2}{2m}\right)
\sim c_2 \ln r$$ for some (computable) constant $c_2$ and all large 
$r$, we get a useful upper bound for any $c_1>0$ when $\alpha<2$. 
If $\alpha=2$, i.e., the repulsive part $U(r)$ decays like a Hardy 
type potential, we also need  that $c_1$ is large enough.

Of particular importance are  multi--particle systems, such as 
$N$ electron atoms with a nucleus of charge $Z$. 
For such atomic systems ground states exist once $N<Z+1$, due to a classical result by Zhislin \cite{Zhi60}. For $N > 2Z+1$, no such states exist \cite{Lie84}.
Hence, for any fixed number $N$ there is a critical charge $Z_c(N)$ such that for $Z > Z_c(N)$ bound states exist and for charges $Z < Z_c(N)$ the quantum system has no bound state. 
Note that $Z_c(N)$ does not have to be a whole number. 

For helium-like systems, a variational calculation of Bethe\cite{Bet29} shows that $Z_c(2)<1$. Numerically, it is known \cite{BakFreDavHil} that $Z_c(2)\sim 0.91$.
The existence and absence of an eigenstate for the simplest nontrivial example of helium-like systems for $Z=Z_c(2)$, was studied extensively by M.\ and T.\ Hoffmann-Ostenhof and Simon \cite{HofOstHofOstSim83}. They derived the existence of an eigenstate at critical coupling $Z_c(2)$ for a singlet state and conjectured its fall-off behavior to be subexponential \cite{HofOstHofOstSim83}. This conjectured fall-off behavior of threshold eigenstates was used for example in \cite{Bur21,GB2015,Hog98,MirUmrMorGor14}. 
Using our method we recently proved in \cite{HunJexLan21-Helium} that the conjecture made in \cite{HofOstHofOstSim83} is correct.

For general atoms, the existence of a ground state at critical coupling was studied in the  Born-Oppenheimer approximation in \cite{BelFraLieSei14} and without it under the additional condition $Z_c(N)\in(N-2,N-1)$ in \cite{GriGar07}. These results establish the existence  
of an eigenstate, but the derived decay bounds are far from what is physically expected~\cite{Hog98}.

Our approach relies mostly on energy estimates which, when combined 
with a geometrically inspired lower bounds for the 
multiparticle potentials of atomic systems, see e.g. \cite{Uch69}, are also applicable to many-particle systems. In particular, our method is applicable to atomic systems under the additional assumption that $N-K>Z_c(N)$, where $K$ is the number of electrons leaving the atom as $Z$ decreases below $Z_c(N)$. A preprint with a proof of concept is available on the arXiv \cite{HunJexLan19-NAtom}.

For very large atoms, it is undoubtedly necessary to use, at least for the inner electrons, the corresponding relativistic equations to obtain the correct results. Our method relies mainly on the IMS localization formula. 
Thus using known results for pseudo-relativistic quantum systems \cite{BarHarHunSem19}, it should be possible to adapt our method to systems with pseudo-relativistic electrons.
Moreover, calculations suggest that our method is also valid within Hartree-Fock and Density Functional Theory (DFT). This is especially interesting due to the fact that these theories are inherently nonlinear.

\bigskip
\noindent
\textbf{Acknowledgements:} 
Funded by the Deutsche Forschungsgemeinschaft (DFG, German Research Foundation) -- Project-ID 258734477 -- SFB 1173 (Dirk Hundertmark). 
This project has received funding from the 
European Research Council (ERC) under the 
European Union's Horizon 2020 research and 
innovation programme (grant agreement MDFT No.\ 725528) (Michal Jex).
Michal Jex also received financial support from the 
Ministry of Education, Youth and Sport of the Czech Republic under the Grant No.\ RVO 14000. This project has received funding from the European Research Council (ERC) under the European Union’s Horizon 2020 research and innovation programme (ERC StG MaMBoQ, grant agreement No.\ 802901) (Markus Lange).


\end{document}